\documentclass[preprint, cite]{epl}

\usepackage{amsmath}
\usepackage{amssymb}
\usepackage{graphicx}
\usepackage{dcolumn}
\usepackage{bm}

\def\ccc#1;#2{\left\langle #1 \left\vert #2 \right.\right\rangle}
\def\ev #1{\left\langle #1 \right\rangle}

\title{Multiscaling and non-universality in fluctuations of driven complex systems}

\author{Zolt\'an Eisler\inst{1}\thanks{Email: \email{eisler@maxwell.phy.bme.hu}} \and J\'anos Kert\'esz\inst{1,2} \and Soon-Hyung Yook\inst{3} \and Albert-L\'aszl\'o Barab\'asi\inst{3}}
\shortauthor{Z. Eisler \etal}
\institute{
  \inst{1} Department of Theoretical Physics, Budapest University of
  Technology and Economics - Budapest, H-1111\\
  \inst{2} Laboratory of Computational Engineering,
  Helsinki University of Technology - Espoo, Finland\\
  \inst{3} Center for Complex Network Research and Department of Physics -
University of Notre Dame, Notre Dame, IN 46556}


\begin{document}

\maketitle

\begin{abstract}
  For many externally driven complex systems neither the noisy
  driving force, nor the internal dynamics are \emph{a priori}
  known. Here we focus on systems for which the time dependent
  activity of a large number of components can be
  monitored, allowing us  to separate each signal into a component attributed to the external driving
  force and one to the internal dynamics. We propose a
  formalism to capture the potential multiscaling in the
  fluctuations and apply it to the high frequency trading
  records of the New York Stock Exchange. We find that on the
  time scale of minutes the dynamics is governed by internal
  processes, while on a daily or longer scale the external
  factors dominate. This transition from internal to
  external dynamics induces systematic changes in the scaling
  exponents, offering direct evidence of non-universality in the
  system.

\end{abstract}

 While it is hard to find a generally accepted definition of complex systems,
most systems that are colloquially labeled "complex" include a
large number of interacting constituents (or nodes) whose
collective dynamics leads to emergent spatial and/or temporal
structures \cite{complexity}. The most studied examples
that fit this paradigm include the cell, vehicular traffic
or the World-Wide-Web \cite{uml}. Very
often these systems operate far from equilibrium and under the
influence of an external driving force. Yet, typically the mechanisms governing the internal
dynamics of these systems are not \emph {a priori} known and even the driving force
is not necessarily under the observer's control.  Moreover, the
separation of the system from its environment is often arbitrary,
making difficult to systematically distinguish the internal from
the external degrees of freedom.

 With the improvement of the measurement and information processing tools
 an increasing number of systems can
be monitored through multichannel measurements, offering the
possibility to record and characterize the simultaneous time
dependent behavior of many of the system's constituents. These
advances in observational techniques offer an important scientific
challenge: Can we design systematic methods that, taking advantage
of the new datasets, can help us to map out the interactions and the
dynamics of various complex systems? Considering the large number
of constituents and the complexity of the behavior displayed by
them the above task is a truly ambitious undertaking, thus even
partial progress is of major potential significance.

Concepts like scaling, multiscaling and universality
\cite{kadanoff} have been found extremely useful in the
characterization of  complex phenomena, as they offer general
relationships, leading to organizing and systematic categorization
principles. Indeed, recent measurements focusing on  the
fluctuations at the ``nodes'' of several complex systems \cite{barabasi.fluct}
indicate that the relationship between the
standard deviation $\sigma_i $ and time average $\ev{f_i}$ of the
$f_i(t)$ signal capturing the time dependent activity of node
$i=1,\dots,N$ follows the scaling law
\begin{equation}
\sigma \propto \langle f \rangle ^\alpha . \label{eq:power-law}
\end{equation}
Yet, this finding leaves a number of important questions
unanswered. For example, the measurements indicate that real
systems belong to one of the two extreme universality classes
characterized by either
  $\alpha \approx 1/2$ (observed for the
Internet and computer chip) or  $\alpha \approx 1$ (highway
traffic, river network and World-Wide-Web), the former describing
an endogenous, while the latter an exogenous dominance in the
fluctuations. Finding systems that display internal organizing
principles  leading to different exponents from these two extreme
values would significantly enrich our understanding of complex
 dynamics. Furthermore, universality in statistical physics
is more than mere numerical agreement of exponents: In critical
phenomena \cite{stanley.critical} whole scaling functions are
expected to be universal. A natural question in this context can
be formulated as follows: Are the distributions of the
fluctuations characterized by universal scaling functions? In this
sense a signature of non-universality would be if some systems
showed multifractality while others did not. Indeed, in the
systems investigated thus far multiscaling appeared to be absent
\cite{menezes.unpublished}. A major goal of this paper is to
introduce the computational tools to uncover  potential
multiscaling in real systems. Finally, we apply these tools to the
stock market, offering direct evidence of both non-universal
exponents and multiscaling.

To capture the  dynamics of systems that can be monitored through
several channels we decompose the signals into two components, that we will call the external and the
internal part \cite{barabasi.separating}. For this we define the
system's \emph{global activity} $F(t)$ as a sum over the activity
of all elements
\begin{equation}
F(t)= \sum_i f_i(t).
\label{eq:Ft}
\end{equation}
As $F(t)$ characterizes the common trends in the systems's
activity, in (\ref{eq:Ft}) the individual, independent
fluctuations of the components are averaged out. The components
are expected to follow this "external" or averaged trend (which itself may be
noisy), while the fluctuations around $F(t)$ are of "internal"
origin, driven by interactions among the system's components. For
a wide variety of cases the total activity $f_i(t)$ of node $i$
can be split into an \emph{external activity} defined as
\begin{equation}
f_i^{ext} (t)= \frac{\ev{f_i}}{\ev{F}} F(t) ,
\label {eq:ext}
\end{equation}
representing node $i$'s expected share  of the global activity
$F(t)$, and the deviations or \emph{internal activity} is given by
\begin{equation}
f_i^{int}(t)= f_i(t)-\frac{\ev{f_i}}{\ev{F}} F(t),
\label {eq:sepa}
\end{equation}
where $\ev{\cdot}$ denotes temporal average \cite{foot1}, while \eqref{eq:sepa} the fluctuating component of a node's activity. By definition, $\ev{f_i^{int}} = 0$. However, \eqref{eq:ext} contains the expected changes in the node's activity, given the overall changes in the system's activity. If the system is closed, then $F(t)$ is time independent. Thus there are no changes in the external component $f_i^{ext}(t)$ either. 

The standard deviation $\sigma^{(\ell)}_i$
of the activity of the $i$th component is
\begin{equation}
\sigma^{(\ell)}_i = \sqrt{\ev{\left ( f_i^{(\ell )} - \ev{f_i^{(\ell
          )}}\right )^2}},
\end{equation}
where the label $(\ell)$ represents \emph{tot}, \emph{int} or
\emph{ext}, indicating that $\sigma_i$ is calculated from the
total, internal or external signal, respectively. We will
everywhere omit $(\ell )$ for the total case.

In order to investigate the multiscaling behavior of the
fluctuations of total noise we propose the multiscaling relation:
\begin{equation}
\label{eq:alphaq}
\ev{\left \vert f_i - \ev{f_i}
  \right \vert^q} = C_F^q (\Delta t, q)\left\langle f_i \right\rangle^{q\alpha (q)}.
\end{equation}
This means that all $q$th order central moments of activity, which characterize fluctuations around the mean
behavior, scale as power-laws with the mean total activity of the
same element. In this notation, $\alpha = \alpha(2)$. Note that analogous definitions can
be given for the internal and the external case. In particular, as
$\ev{\left \vert f_i^{ext}\right \vert^q} \equiv {\ev{f_i}^q},$
for external activity we have $\alpha (q;ext)\equiv 1$ [see \eqref{eq:ext}], indicating that multiscaling can be
present only in the internal or total fluctuations. In
\eqref{eq:alphaq} $\alpha (q)$ is a formal analogue to the
generalized dimensions $H(q)$ for multifractal or multiaffine time
series \cite{vicsek.book}, defined as
\begin{equation}
\ev{\left \vert f_i - \ev{f_i}\right \vert^q} =
C_T^q(i, q) \Delta t^{qH(i, q)},
\label{eq:multifractal}
\end{equation}
where $\Delta t$ is the time above which the averages are to be taken.
Combining \eqref{eq:alphaq} and \eqref{eq:multifractal} one can eliminate a variable,
obtaining:
\begin{eqnarray}
\frac{\left\langle f_{i;\Delta t} \right\rangle^{\alpha (\Delta t, q)}}{\left\langle f_{i;\Delta t} \right\rangle^{\alpha (\Delta t, q')H(i, q)/H(i, q')}}\frac{C_F (\Delta t, q)}{C_F (\Delta t, q')^{H(i, q)/H(i, q')}} = \frac{C_T(i, q)}{C_T(i, q')^{H(i, q)/H(i, q')}}.
\label{eq:corresp5}
\end{eqnarray}

Here the r.h.s. does not depend on $\Delta t$, thus the l.h.s. should not
either. Information on the temporal scaling goes into the r.h.s. in
form of the exponent ratios  $H(q)/H(q')$. However, if the time series
does not show multifractality, i.e., $H(q)=H(q')$ then the simple
moment ratios from \eqref{eq:alphaq} should not depend on the the averaging time:

\begin{eqnarray}
\frac{C_F (\Delta t, q)\left\langle f_{i;\Delta t} \right\rangle^{\alpha (\Delta t, q)}}{C_F (\Delta t, q')\left\langle f_{i;\Delta t} \right\rangle^{\alpha (\Delta t, q')}}= \frac{C_T(i, q)}{C_T(i, q')}.
\label{eq:unite4}
\end{eqnarray}

Let us turn to an example where the tools
introduced above prove useful.
In the study of the financial market as a complex system statistical
physics concepts turned out to be very powerful \cite{stanley.bouchaud.book}.
Enormous amount of data is available as every
transaction is recorded on the stock market. A
natural choice of constituents here are the stocks of firms that are
publicly traded. In order to keep the analogy with previously
studied cases we have chosen the flow as the
signals. On a given time horizon $\Delta t$, let the (total) \emph{activity}
or flow
of the $i$th stock at time $t$ be
\begin{equation}
f_i^{\Delta t}(t) = \sum_{\tau \in [t, t+\Delta t]} Y_i(\tau ) V_i(\tau ),
\label{eq:flow}
\end{equation}
where $\tau$ runs for all trades of the $i$th stock in the given
interval. This corresponds to the coarse-graining of the individual
events, or the so-called tick-by-tick data. $ Y_i(\tau )$ is the price
and $V_i(\tau)$ is the traded volume for the trade at time
$\tau$ \cite{Gopi_volume}. We used data with a minute resolution where
the price within the minute was that of the last trading minute (which
causes a negligible error). Hence, $f_i^{\Delta t}(t)$ gives the total
traded value of stock $i$ between times $t$ and $t+\Delta t$. $\Delta
t$ can be chosen as any multiple of $1$ minute, and $t = 0, \Delta t,
2\Delta t, \dots$. In the following for simplicity of notation we will
omit $\Delta t$'s where they are kept constant, only indicating their
value once.

For empirical analysis we used the TAQ database \cite{taq} of the
New York Stock Exchange for the period of 2000-2002, which after
some filtering\cite{foot2} contains $N=2200$ stocks.

\begin{figure}[!htb]
\centerline{\hbox{\includegraphics[height=115pt]{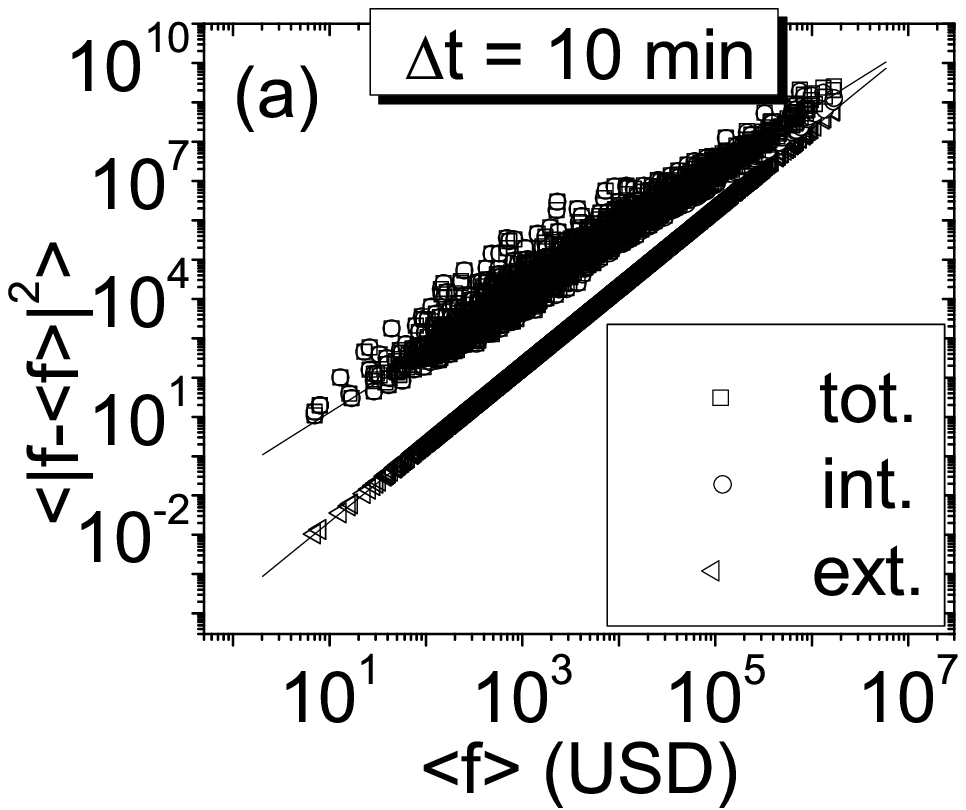}
\hskip30pt\includegraphics[height=115pt]{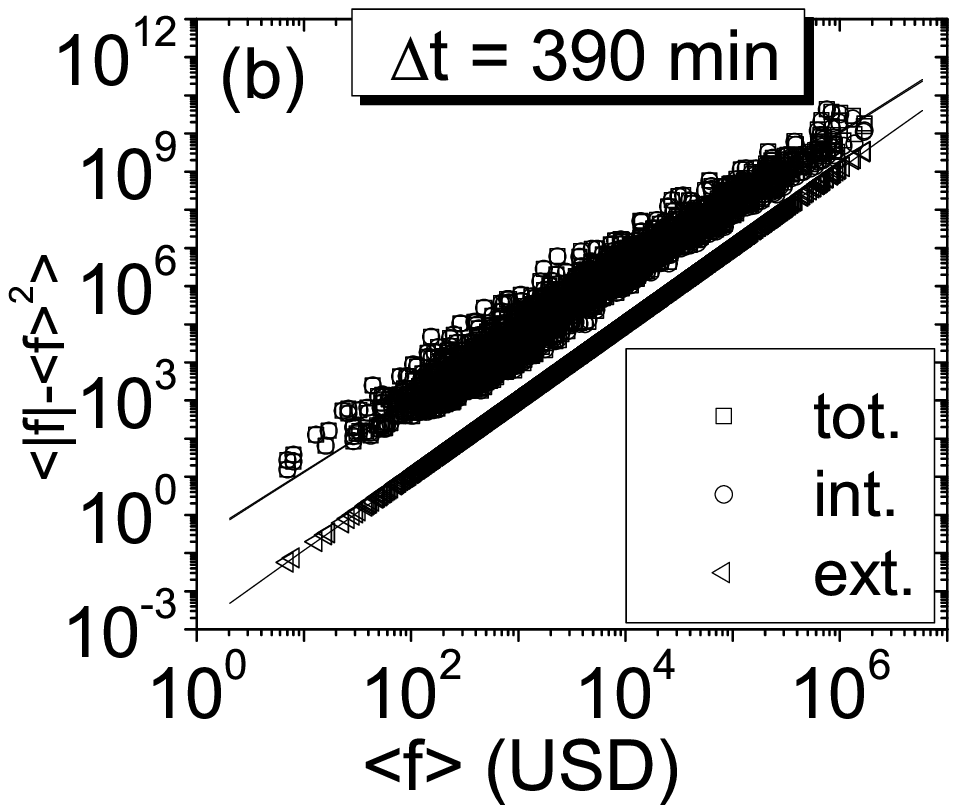}}}
\caption{Scaling of $q=2$ central moments of activities and fitted
linear trends. {\bf (a)} $\Delta t = 10$ minutes. $\alpha (2;tot)
\approx \alpha (2;int) = 0.773 \pm 0.003$. {\bf (b)} $\Delta t = 390$
minutes = $1$ trading day. $\alpha (2;tot) \approx \alpha (2;int) =
0.885 \pm 0.005$. Errors are estimated from the linear fit on the log-log plot. Note that although the errors are very low due to the large number of data points, there is pronounced unexplained variance around the expected scaling of total and internal noise. This can be reduced significantly if one takes the system structure (in our case the clustering of stocks into market sectors) into account \cite{coming}.}
\label{fig:tq2}
\end{figure}

\begin{figure}[!htb]
\vskip2mm
\centerline{\hbox{\includegraphics[height=115pt]{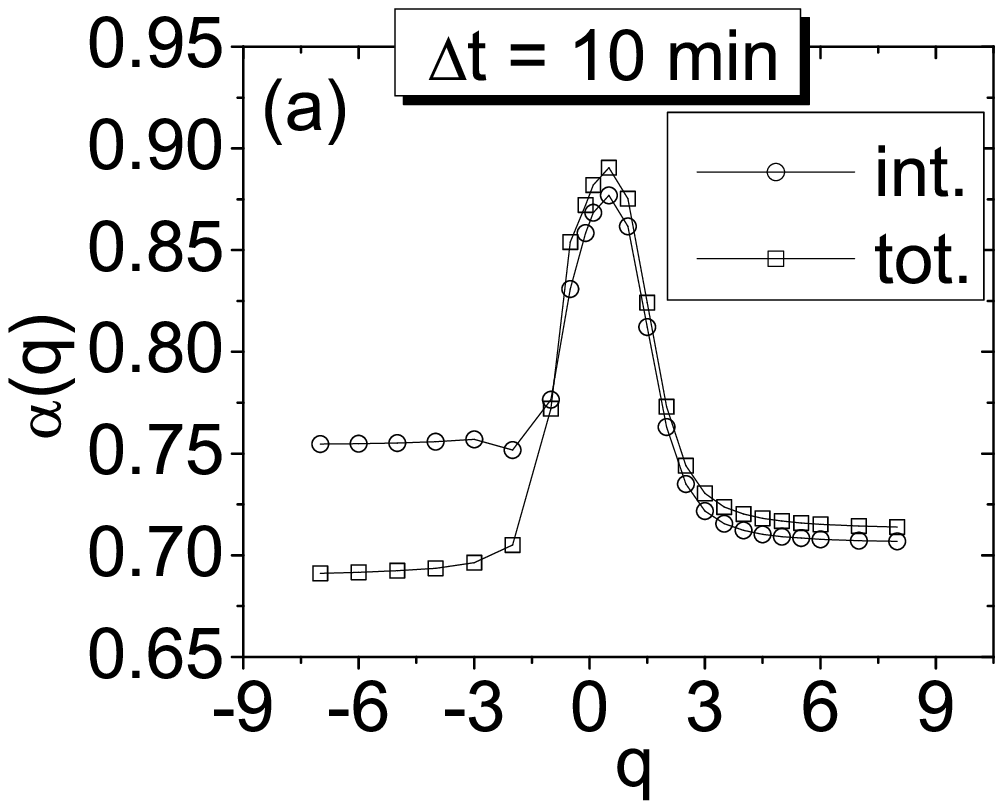}
\hskip10pt\includegraphics[height=115pt]{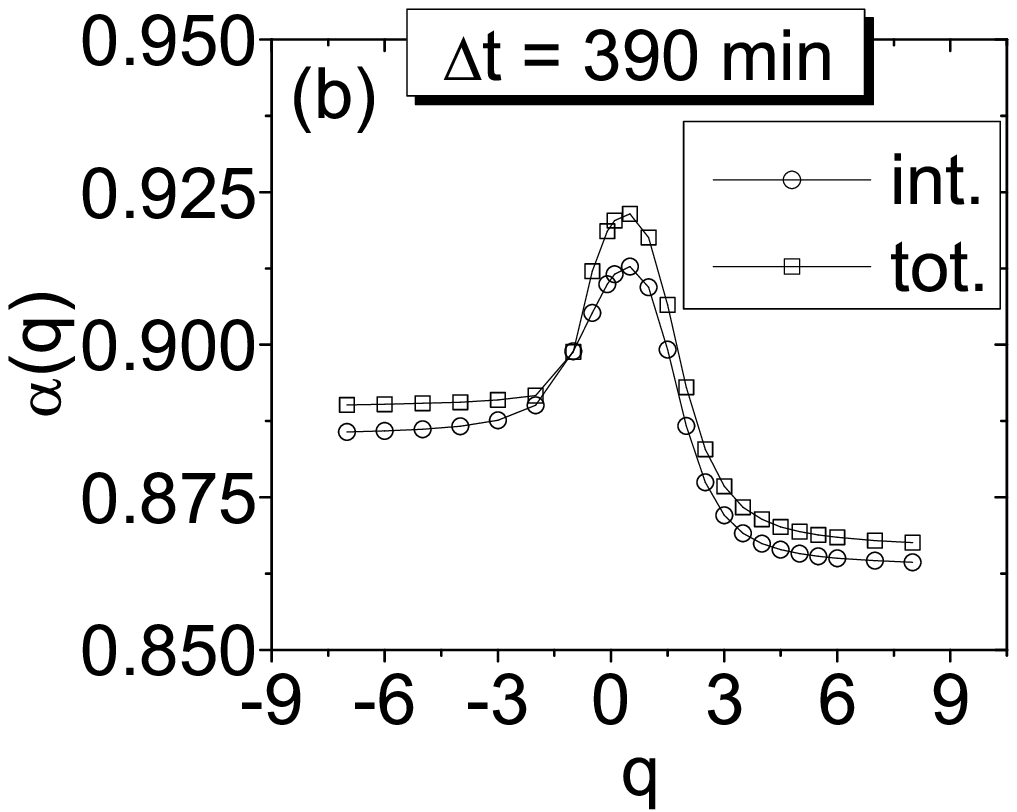}}}
\caption{The $\alpha (q)$ multiscaling exponents for all three types
of activity. $\alpha(q;ext)\equiv 1$. {\bf (a)} $\Delta t = 10$
minutes. {\bf (b)} $\Delta t = 390$ minutes = $1$ trading day.}
\label{fig:alphaq}
\end{figure}

The $\alpha (q)$ scaling exponents were measured for fixed $\Delta
t$, examples being shown in Fig. \ref{fig:tq2}. The first striking
observation is that for the second moment we find power law
behavior over five orders of magnitude with an exponent which is
significantly different from both 1/2 and 1. Second, we find
multiscaling (i.e., a dependence of $\alpha (q)$ on $q$, except
for the trivial case of external activity), as shown in Fig.
\ref{fig:alphaq}. Furthermore, the exponents show a strong
dependence on the time horizon $\Delta t$.


\begin{figure}
\vskip2mm
\centerline{\includegraphics[width=130pt]{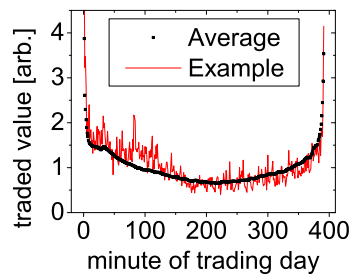}}
\caption{The average intraday pattern of trading
  activity. The activity time series of one example day is given for
  comparison.}
\label{fig:pattern}
\end{figure}

In the activity  \eqref{eq:flow} there are two sources of external
impact, a random and a regular part. The latter is manifest in,
e.g., different kinds of seasonalities and intraday patterns.
Since we are interested in the \emph{fluctuations} it is  natural
to detrend the data from such regular contributions. In this
respect intraday patterns are particularly strong
\cite{stanley.bouchaud.book}: At the beginning and end of the
trading day the activity is anomalously high. In addition, one
finds a small irregularity in activity right after 10 a.m., which
is a typical time for news arrival. The pattern, averaged for all
full-length days in our dataset, is shown in Fig.
\ref{fig:pattern} which is used for detrending, achieved by
dividing traded values by the respective values of
the trend pattern.

\begin{figure}[!htb]
\vskip2mm
\centerline{
\includegraphics[height=115pt]{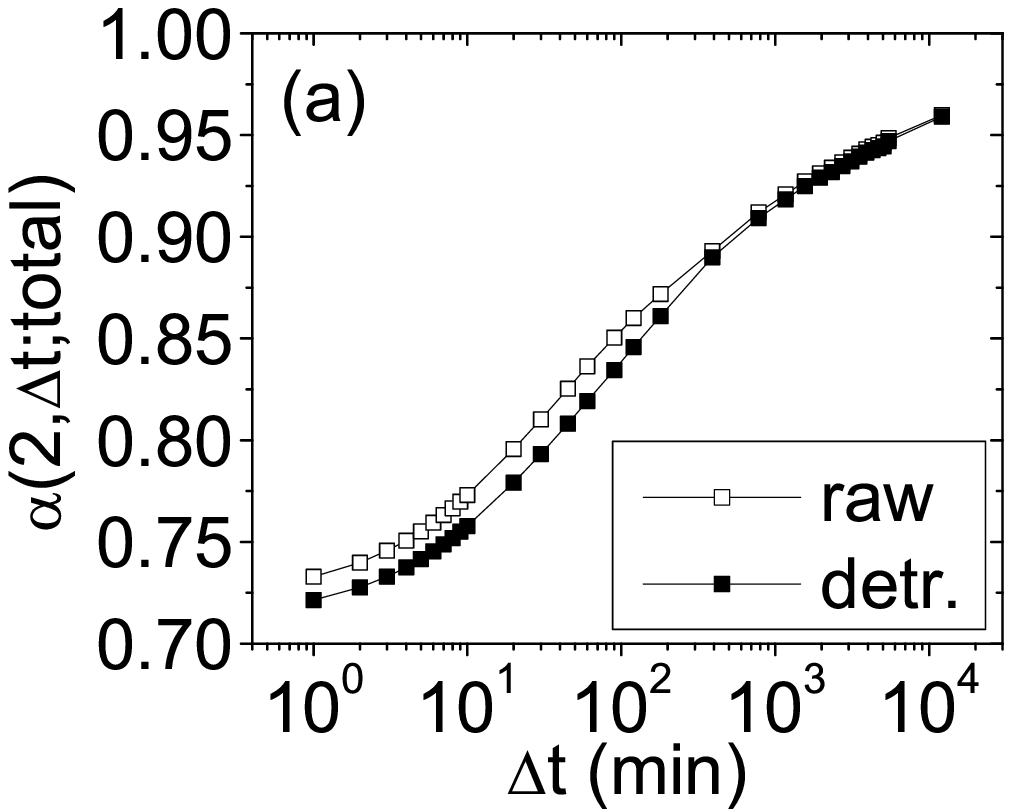}
\includegraphics[height=115pt]{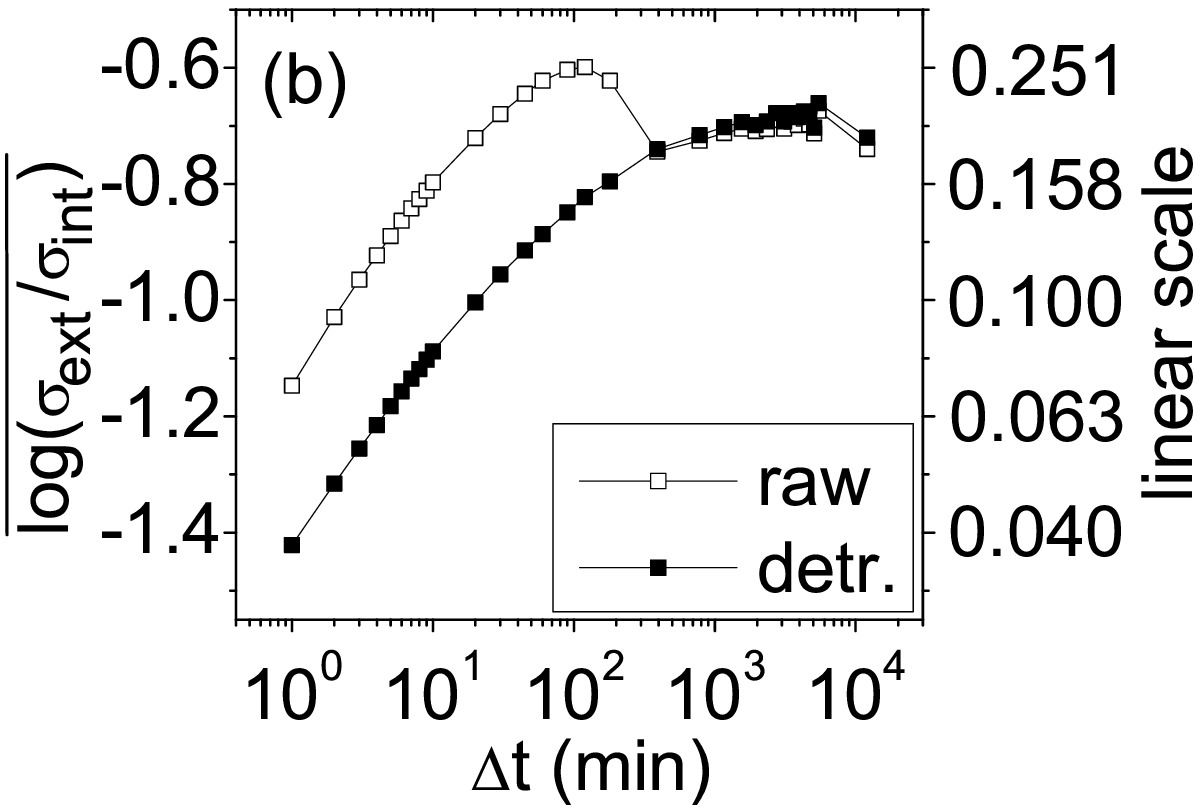}
}
\caption{{\bf (a)} Values of $\alpha$ as a function of the time scale
$\Delta t$. There is a clear tendency present. The exponent
grows from a lower value corresponding to the dominance of internal
dynamics to nearly $1$, which characterizes externally driven
systems. {\bf (b)} The ratio of the standard deviation of fluctuations
originating from external and internal sources. Raw data shows an
anomalous behavior, the result of peaks in trading activity at the
beginning and the end of each day of trading. Results for data
detrended with the average of this daily pattern show an effect
similar to the one in the behavior of the exponent $\alpha$. In the
$\Delta t = 1 \dots 390$ minutes $=1$ day interval, there is a faster
increase in the contribution of external fluctuations. Beyond the
scale of a few days this tendency saturates.}
\label{fig:time_horizons}
\end{figure}

We measured the scaling exponents $\alpha$, using fits similar to
those in Fig. \ref{fig:tq2}. Simultaneously, we calculated the
values of $\log (\eta(\Delta t))$ (with $\eta = \sigma_{ext}
/\sigma_{int}$) averaged over all stocks. Both
results are shown in Fig. \ref{fig:time_horizons}, indicating a significant
$\Delta t$ dependence. One finds that as we move from the minute
scale to the weekly scale, the nature of fluctuations changes
gradually toward the externally driven limit. This is in
correspondence with results for the $\eta$ ratio (see Fig.
\ref{fig:time_horizons}(b)). Detrending of the data causes a little
systematic decrease in $\alpha$ values for time horizons less than
one day.

An anomalous non-monotonicity is present in the raw data [see
Fig. \ref{fig:time_horizons}(b)], while for the detrended data,
we recover the expected tendency: As
$\Delta t$ increases, the external driving force gets stronger.
The range of $\eta $ is similar to that in
\cite{barabasi.fluct}. We find, that for $\Delta t$'s larger than
a few days, the external contribution to fluctuations saturates.

\begin{figure}[!htb]
\vskip2mm
\centerline{\hbox{\includegraphics[height=115pt]{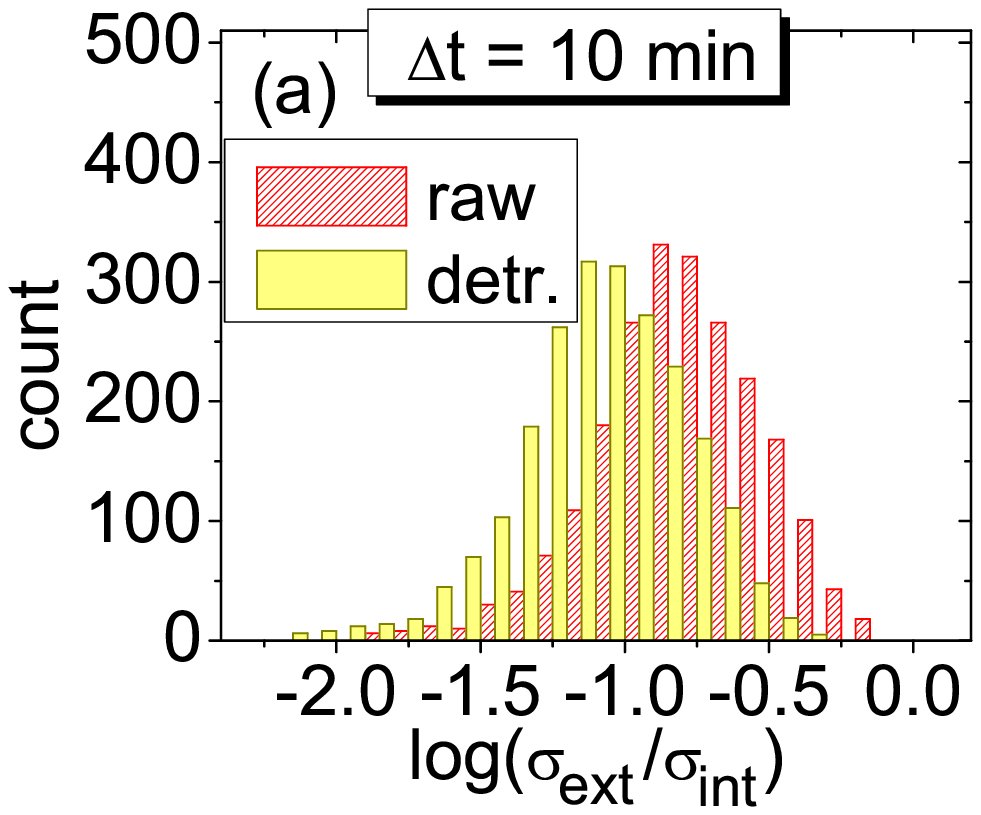}
\includegraphics[height=115pt]{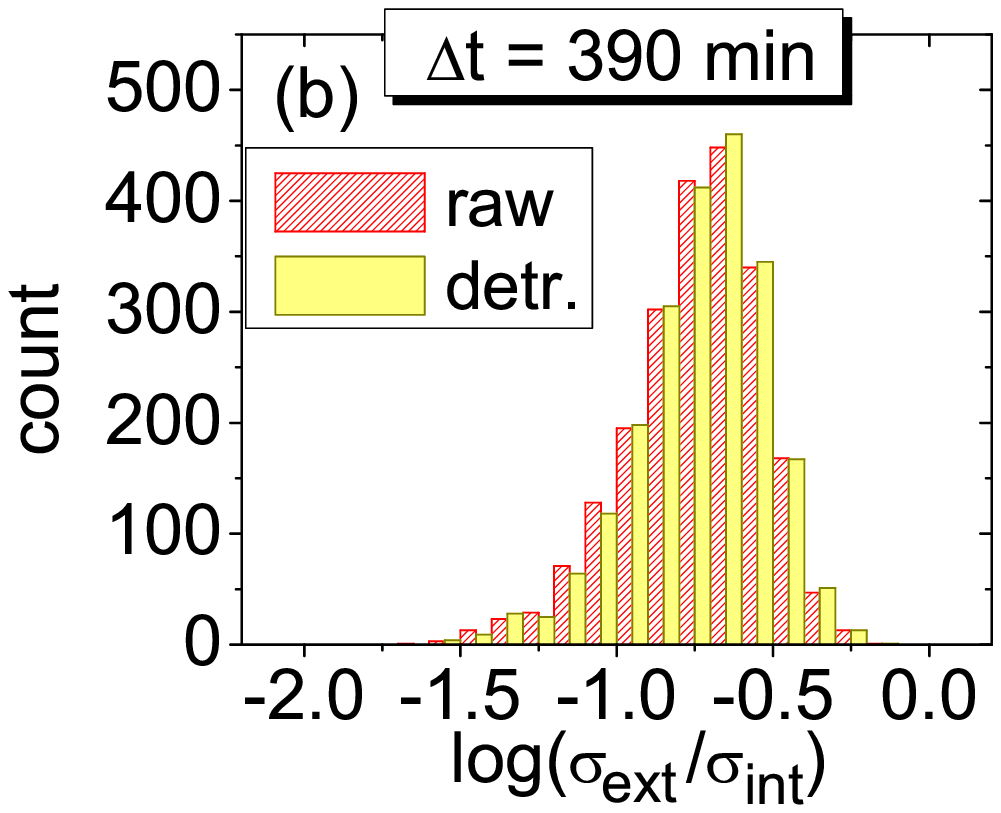}}}
\caption{Histogram of the ratios of the internal and
external standard deviations for all stocks, both for raw and
detrended time series. The comparison shows, that external influence
is stronger on a daily time scale than for fine resolution data. {\bf
(a)} $\Delta t = 10$ minutes. Detrending with the intraday pattern
removes the spurious external driving force due to daily peaks in
activity. This decreases the measured external contribution in
fluctuations. {\bf (b)} $\Delta t = 390$ minutes = $1$ trading
day. Detrending does not change behavior measured on a daily (or
longer) scale.}
\label{fig:logratio}
\end{figure}

As we change the time scales ($\Delta t = 1$ minute$\dots 2$ weeks)
we observe a transition
between two limits. With decreasing $\Delta t$, we get closer to the
limiting endogenous behavior $\alpha = \alpha_{int} < 1$, and $\eta \ll 1$.
Yet, Fig. \ref{fig:time_horizons}(a) indicates, that even the
limiting value $\alpha^*_{int}$ should significantly differ from
$1/2$ measured in several systems\cite{barabasi.fluct, foot3}.
This suggests that despite earlier observations
\cite{barabasi.fluct} the behavior $\alpha^*_{int}=1/2$ is
not universal, although the mechanism responsible for this
non-trivial scaling is unknown. Furthermore, we observe a
continuous dependence of the exponent $\alpha_{int}$ on $\Delta t$
with scaling spanning over five orders of magnitude, indicating
that the observed exponents are likely not due to finite size
crossover phenomena.  With increasing $\Delta t$ there is a
growing role of external forces, and beyond the daily scale we
reach the exogenous limit, where $\alpha \rightarrow 1$ and $\eta
\sim 0.20$. In this limit the dynamics is dominated by the
external  driving force. This mechanism is behind the so called
Epps-effect \cite{epps.mantegna}, namely that in high frequency
data the cross correlations between stock data are much less
pronounced than in, say, daily ones: Correlations reflect the
similarities in how different stocks react to external effects and
this is covered for short time horizons by noisy internal
dynamics.

These results offer a coherent qualitative picture about market
dynamics. The impact of incoming news needs a finite time to diffuse.
Hence, on short time scales, the response to them is small. The factor
that determines the fluctuations of trading activity is internal: it
is the trading mechanism itself. On daily or longer scales, however,
the internal fluctuations have smaller importance, and the market
tends to move with the global activity. In periods of ``business as
usual'' the natural human scale of one day seems to be needed to reach
a kind of coherence: News and trends can be evaluated, informations
are exchanged and collective decisions are made. Interestingly, the
scaling of asset return distributions \cite{stanley.bouchaud.book}
also breaks down on the scale of one day, see e.g. \cite{foot4}.

Distinguishing between endogenous and exogenous origins of market
events is a central research problem \cite{cutler.sornette.zawa}.
Though theoretically and in agent-based model calculations it has
been possible to investigate this question, its empirical study is
extremely difficult.  The appealing feature of the presented
method is that it is based purely on multichannel time series and
no knowledge of the internal structure or the dynamics goes into
it. Thus it can serve as an empirical foundation for simulations
and further theoretical work. Clearly, its applicability goes much
beyond the examples discussed so far.

In summary, we have introduced a multiscaling formalism to study
fluctuations in complex systems. We find that non-universal
behavior is manifested not only in $\alpha $ exponents different
from the universal values 1/2 and 1 but also in the scaling
properties of the distribution functions. The $\alpha $ exponents
found for the flow data of 2,200 stocks on the NYSE showed a
continuous dependence on the time horizon with good quality
scaling over five orders of magnitude. An additional signature of
non-universal behavior is that multiscaling was found, in contrast
to several other complex systems investigated in the similar
fashion \cite{menezes.unpublished}.

Acknowledgments: We thank Marcio de Menezes for useful comments
and  Gy\"orgy Andor for his help with the data. JK thanks the
University of Notre Dame for hospitality, and he is member of the
Center for Applied Mathematics and Computational Physics, BUTE.
Research at Notre Dame was supported by NSF.


\begin{thebibliography}{99}
\bibitem{foot1}
We note that this separation does not lead to uncorrelated
decomposition which can be achieved by projector technique \cite{coming}.
\bibitem{foot2}
We discarded all companies, that were not
continuously present on the market over the $3$-year interval and we
combined different class stocks of the same company.
\end{thebibliography}
\end{document}